\documentclass[12pt]{article}

\usepackage[utf8]{inputenc}
\usepackage[T1]{fontenc}
\usepackage{mathpazo}
\usepackage[final,expansion=false]{microtype}

\usepackage[letterpaper, margin=1in]{geometry}
\usepackage{setspace}
\usepackage{amsmath, amssymb, amsfonts, amsthm}
\usepackage{mathtools}
\usepackage{bm}

\theoremstyle{plain}
\newtheorem{assumption}{Assumption}
\newtheorem{theorem}{Theorem}

\newtheorem{proposition}{Proposition}

\theoremstyle{definition}

\newcommand{\E}{\mathbb{E}} 

\newcommand{\Prb}{\mathbb P}
\newcommand{\R}{\mathbb R}

\newcommand{\aset}{\mathcal H}

\usepackage{graphicx}
\usepackage[font={footnotesize},justification=justified]{caption}
\usepackage{booktabs}
\usepackage{adjustbox}

\usepackage{enumitem}

\usepackage[svgnames]{xcolor}
\usepackage[colorlinks=true, linkcolor=black, citecolor=black,
            urlcolor=blue]{hyperref}
\usepackage{natbib}
\usepackage[capitalize, noabbrev, nameinlink]{cleveref}
\crefname{assumption}{Assumption}{Assumptions}

\usepackage{tikz}
\usetikzlibrary{calc, positioning}
\usepackage{pgfplots}
\pgfplotsset{compat=1.18}
\usepgfplotslibrary{ternary}

\title{Identifying treatment effects for categorical outcomes in IV models\thanks{%
Corresponding address: oboussim@calpoly.edu, Department of Economics, The California Polytechnic State University, San Luis Obispo.}}
\author{Onil \textsc{Boussim}}

\date{\today}

\makeatletter
\ifcsname c@condition\endcsname
\else
\newtheorem{condition}{Condition}
\fi
\makeatother

\begin{document}

\maketitle

\begin{abstract}
\noindent

This paper studies population treatment effects when the outcome is unordered categorical, the treatment is binary, and the instrument is binary. I introduce an assumption called association similarity. For each instrument value, association similarity requires the odds-ratio association between potential treatment and potential outcome to be the same whether the outcome is evaluated under treatment or under no treatment. Under association similarity and a testable categorywise dominance condition, the full potential-outcome distributions are point identified. The identification is constructive and allows estimation using sample frequencies. I also consider weaker restrictions on departures from association similarity that lead to sharp partial identification. I illustrate the method with an application about the effect of insurance on health outcomes.\\


\end{abstract}

\newpage

\section{Introduction}

A central challenge in causal inference is endogeneity. Many treatments of interest result from choices made by agents who act on private information that also affects the outcomes the researcher seeks to explain. Individuals decide whether to enroll in education programs, firms choose whether to adopt new technologies, and governments design tax or subsidy schemes. These decisions are typically correlated with unobserved factors such as motivation, ability, or local market conditions, which simultaneously influence outcomes of interest. Instrumental variables methods provide a classical solution by using exogenous variation in the probability of treatment to separate causal effects from endogenous selection.

Traditional instrumental variables frameworks are designed primarily for scalar or continuous outcomes, such as income or test scores. Many policy-relevant outcomes, however, are inherently categorical and unordered: employment sector, educational attainment, product choice, voting behavior, or self-reported health. In such settings, researchers are often interested not only in an average effect but in how treatment shifts the entire distribution of outcomes across mutually exclusive categories. Despite its empirical relevance, IV analysis for unordered categorical outcomes remains underdeveloped. Existing methods often reduce the outcome to a binary indicator, thereby discarding distributional information, or impose parametric restrictions such as multinomial logit or nested logit models. Other approaches are tied to ordered or continuous outcomes and do not extend naturally to unordered settings.

This paper develops a nonparametric framework for identifying and estimating the entire distribution of potential outcomes when the observed outcome takes values in a finite set of unordered categories. The main identification result relies on a new assumption I call \emph{association similarity}. This assumption states that the way unobserved characteristics relate to potential outcomes is the same whether an individual is treated or not.

To illustrate, suppose the instrument is an encouragement to enroll in a job training program, and the outcome is employment in one of several sectors, such as manufacturing, services, or unemployment. Unobserved traits such as motivation, ability, or access to transportation may affect both whether someone enrolls when encouraged and which sector they would work in. Association similarity requires that the link between these unobserved traits and sectoral employment does not depend on treatment status.

More formally, I use the discrete dependence idea of \cite{geenens2020}. In practice, association similarity requires the odds-ratio association between the potential treatment decision and the potential outcome to be the same under treatment and no treatment. These associations may differ across instrument values and across categories. The restriction is formulated directly on finite probability tables. It neither orders the categories nor assumes a multinomial response model.

Under association similarity, the selection bias affecting observed outcomes is the same across treatment states. This makes it possible to cancel it out using variation in the instrument. The result is a set of simple nonparametric formulas that uniquely recover the full distribution of potential outcomes for each category. Recognizing that association similarity may be too strong in some applications, I also propose a relaxation that yields sharp partial identification bounds. These bounds nest point identification as a limiting case and facilitate transparent sensitivity analysis.

I illustrate the method by studying the effect of private insurance on health outcomes. I consider physical and mental health with categories (excellent, good, poor). The instrument is constructed using firm size. I find mixed evidence of the effect of private insurance on health outcomes.

\section{Related Literature}
\label{sec2}

Identifying causal effects with categorical outcomes poses distinct challenges compared with the scalar case (see \cite{graves2022difference}, \cite{boussim2025compositional}, \cite{boussim2026compositional}). While IV methods for continuous outcomes are now well established; see the survey by \citealt{mogstad2024instrumental}, progress for discrete outcomes has been slower.

Early contributions focused on parametric models, such as multinomial probit and logit models with correlated errors \citep{van1981demand,dubin1984econometric,dubin1989selection}. These models achieve identification only under strong distributional assumptions. Related work on bivariate probit models includes \cite{freedman2010endogeneity}, \cite{mourifie2014note}, \cite{han2017identification,han2019estimation}, and \cite{acerenza2023testing}. These approaches remain largely parametric in nature.

Nonparametric approaches offer alternatives that relax these restrictions. \cite{angrist1995identification} provided an early contribution by showing identification of local average treatment effect models. However, that framework delivers the potential outcome distribution only for compliers rather than for the full population. More recently, \cite{chernozhukov2023distribution,chernozhukov2024estimating} developed a local Gaussian representation that, under a copula invariance assumption, enables nonparametric identification in selection models. Their strategy, however, is restricted to \emph{ordered} categorical outcomes because it exploits cumulative distribution functions. The present paper adapts and extends that logic to unordered categorical outcomes by working with a multinomial log-odds parametrization, which does not require any ordering.

A distinct strand of literature adopts a partial identification approach. \cite{manski1990nonparametric} derived the sharpest bounds on treatment effects consistent with the data and basic assumptions. \cite{balke1997bounds} obtained sharp IV bounds for binary outcomes under no defiers. \cite{vytlacil2007dummy} and \cite{shaikh2011partial} extended partial identification results to more general limited dependent variable settings. \cite{beresteanu2021quantile} and \cite{chesher2010instrumental} studied sharp identification regions using random set theory and quantile methods, respectively. The present paper contributes to this literature by providing sharp bounds under weakened forms of association similarity. The framework also relates to the sensitivity analysis literature initiated by \cite{rosenbaum1987sensitivity} and developed in the IV context by \cite{conley2012plausibly} and \cite{nevo2012identification}, insofar as bounded association serves as a formal sensitivity parameter.

\section{Framework and discrete dependence}

\subsection{Potential outcomes and latent probability tables}

I adopt the potential outcomes framework to formalize causal inference with instrumental variables when the outcome variable is categorical. Let the treatment be binary, $D \in \{0,1\}$, and let the instrumental variable also be binary, $Z \in \{0,1\}$. The outcome of interest, $Y$, takes values in a finite set of $q$ mutually exclusive and exhaustive unordered categories:
\[
Y \in \mathcal{C} = \{c_1, c_2, \dots, c_q\}.
\]
Because the outcome is unordered, cumulative distribution functions and ordered-threshold models are not applicable. One must work directly with the vector of category probabilities.

For each individual, define potential outcomes $Y_0$ and $Y_1$, where $Y_d$ represents the outcome that would be observed under treatment status $D=d$. Each $Y_d$ takes values in $\mathcal{C}$. The objects of interest are the full distributions of potential outcomes. For each $d \in \{0,1\}$, define
\[
\pi_d
=
\bigl(\Prb(Y_d=c_1),\dots,\Prb(Y_d=c_q)\bigr)^\top
\in \mathcal{S}^{q-1},
\]
where $\mathcal{S}^{q-1}$ denotes the $(q-1)$-dimensional probability simplex.

To characterize treatment selection, I introduce potential treatment indicators $D_0$ and $D_1$, where $D_z \in \{0,1\}$ denotes the treatment decision the individual would make under instrument value $Z=z$. The observed treatment and outcome satisfy
\[
D = ZD_1 + (1-Z)D_0,
\qquad
Y = DY_1 + (1-D)Y_0.
\]
The target parameters are
\begin{equation}\label{eq:target}
\pi_{dk}=\Prb(Y_d=c_k),
\qquad
\tau_k=\pi_{1k}-\pi_{0k},
\qquad
\sum_k \tau_k=0.
\end{equation}

\subsection{Discrete dependence: the nucleus of a probability table}

A central difficulty in extending copula ideas to discrete outcomes is deciding what should be regarded as dependence independently of the marginal distributions. For continuous random variables, Sklar's theorem provides a familiar answer: the joint distribution can be decomposed into marginal distributions and a copula, and the copula is invariant to strictly increasing transformations of either variable. This separation is useful because it allows one to compare dependence patterns without being affected by changes in the marginal distributions.

This argument, however, cannot be transferred directly to discrete random variables. The probability integral transform is no longer continuously uniform, and a discrete joint distribution is generally compatible with multiple classical copulas. Consequently, the copula is not uniquely determined by the joint distribution. The usual construction is also inherently tied to ordered variables and is therefore less natural for unordered categorical outcomes.

Following \citet{geenens2020}, I instead define dependence through the transformations that leave it unchanged. Consider a bivariate probability table. Changing its row or column margins should not, by itself, alter the intrinsic association between the two variables. Thus, two probability tables should represent the same dependence structure whenever one can be obtained from the other by separately reweighting rows and columns.

More formally, let $\mathcal{P}_{R\times S}$ denote the set of strictly positive $R\times S$ probability tables $\mathbf p=(p_{xy})$, where $x\in\{0,\ldots,R-1\}$ and $y\in\{0,\ldots,S-1\}$. Two tables belong to the same dependence class if they differ only through row and column rescaling. This operation is often referred to as \emph{matrix scaling} or \emph{marginal distortion}: it changes the marginal distributions while preserving the association encoded in the table. The key invariants under such transformations are the odds ratios. Using cell $(0,0)$ as a reference, define
\begin{equation}
\label{eq:oddsratio}
\omega_{xy}(\mathbf p)
\equiv
\frac{p_{00}p_{xy}}{p_{0y}p_{x0}},
\qquad
x=1,\ldots,R-1,
\quad
y=1,\ldots,S-1.
\end{equation}
Collecting these quantities yields the $(R-1)\times(S-1)$ odds-ratio matrix
\[
\Omega(\mathbf p)
\equiv
\left[
\omega_{xy}(\mathbf p)
\right]_{x=1,\ldots,R-1;\ y=1,\ldots,S-1}.
\]

The odds-ratio matrix provides exactly the information about association that remains once the margins are allowed to vary. Multiplying rows and columns by arbitrary positive constants leaves every $\omega_{xy}$ unchanged. Conversely, under positivity, two probability tables with the same odds-ratio matrix differ only by such a marginal distortion. Thus, the odds ratios characterize the dependence structure independently of the marginal distributions.

\section{Identification results}

\subsection{Point identification}
I begin with the classical assumptions for instrument validity.

\begin{assumption}[Instrument validity]
\label{a:iv}
$(Y_0,Y_1,D_0,D_1)$ is independent of $Z$, and $0<\Prb(Z=1)<1$. Outcomes satisfy consistency and exclusion as specified above.
\end{assumption}

Assumption~\ref{a:iv} means that the instrument is independent of all potential outcomes and all potential treatment decisions. It implies that the instrument cannot directly shift the distribution of $Y_d$, that there are no unobserved confounders jointly affecting $Z$ and $(Y_0,Y_1)$, and that individuals do not self-select into instrument values in a way correlated with potential treatment decisions.

To formulate a restriction on selection, consider the latent joint distributions of potential outcomes and potential treatment decisions:
\begin{equation}\label{eq:potentialtable}
q^d_{z,k,j}=\Prb(Y_d=c_k,D_z=j),
\qquad
\Pi^d_z=\bigl(q^d_{z,k,j}\bigr)_{k=1,\ldots,q;\ j=0,1}.
\end{equation}
The $q\times 2$ table $\Pi^d_z$ describes the joint distribution of potential outcome $Y_d$ and potential treatment decision $D_z$. Its row margin is $\pi_d$, and its column margin is $(1-p_z,p_z)$, where
\[
p_z=\Prb(D_z=1).
\]
At a fixed $z$, the treatment margins of $\Pi^0_z$ and $\Pi^1_z$ are identical, but their outcome margins can differ.

\begin{assumption}[Interior tables]
\label{a:positive}
For every $d,z,k,j$, $q^d_{z,k,j}>0$.
\end{assumption}
Assumption~\ref{a:positive} ensures $0<p_z<1$ and makes it possible to define odds ratios. The identification argument proceeds by completing the two potential-outcome tables. The observed data reveal one column of each table. Association similarity links the missing columns to the observed columns, while instrument validity links the tables across instrument values. Together, these restrictions reduce the problem to two unknown numbers, determined by two adding-up equations. Under instrument validity and exclusion, the observed data identify
\[
q^0_{z,k,0}
=
\Pr(Y=c_k,D=0\mid Z=z),
\qquad
q^1_{z,k,1}
=
\Pr(Y=c_k,D=1\mid Z=z).
\]
Thus, for each instrument value, I observe the distribution of $Y_0$ among the untreated and the distribution of $Y_1$ among the treated. The missing probabilities are
\[
q^0_{z,k,1}=\Pr(Y_0=c_k,D_z=1),
\qquad
q^1_{z,k,0}=\Pr(Y_1=c_k,D_z=0).
\]
They describe untreated potential outcomes among those who would receive treatment, and treated potential outcomes among those who would remain untreated. Recovering these probabilities identifies the population margins because
\[
\pi_{0k}=q^0_{z,k,0}+q^0_{z,k,1},
\qquad
\pi_{1k}=q^1_{z,k,0}+q^1_{z,k,1}.
\]
The right-hand sides must give the same margins at $z=0$ and $z=1$. Because the marginal distribution of $Y_0$ is common across instrument values,
\[
q^0_{0,k,0}+q^0_{0,k,1}
=
q^0_{1,k,0}+q^0_{1,k,1}.
\]
Applying the same argument to $Y_1$ gives
\[
q^1_{0,k,0}+q^1_{0,k,1}
=
q^1_{1,k,0}+q^1_{1,k,1}.
\]
Define the observed category-specific differences
\begin{equation}\label{eq:Delta}
\Delta q^0_k
=
q^0_{0,k,0}-q^0_{1,k,0},
\qquad
\Delta q^1_k
=
q^1_{1,k,1}-q^1_{0,k,1}.
\end{equation}
Then the missing cells at $z=1$ can be written in terms of the missing cells at $z=0$:
\begin{equation}\label{eq:dominance}
q^0_{1,k,1}
=
q^0_{0,k,1}+\Delta q^0_k,
\qquad
q^1_{1,k,0}
=
q^1_{0,k,0}-\Delta q^1_k.
\end{equation}
The subtraction order differs deliberately. The first difference measures the decline in untreated category-$k$ mass as the instrument increases; the second measures the increase in treated category-$k$ mass. I now introduce the association similarity assumption:
\begin{assumption}[Association similarity]
\label{a:similar}
For each $z\in\{0,1\}$, the joint distributions of $(Y_0,D_z)$ and $(Y_1,D_z)$ have the same discrete dependence nucleus:
\begin{equation}\label{eq:associationsimilarity}
\Omega(\Pi^0_z)=\Omega(\Pi^1_z).
\end{equation}
\end{assumption}

In explicit potential-outcome probabilities, define
\begin{equation}\label{eq:or}
\omega^d_{zk}
=
\frac{\Prb(Y_d=c_1,D_z=0)\Prb(Y_d=c_k,D_z=1)}
     {\Prb(Y_d=c_1,D_z=1)\Prb(Y_d=c_k,D_z=0)},
\qquad k\geq 2.
\end{equation}
Association similarity is precisely the restriction
\[
\omega^0_{zk}=\omega^1_{zk}
\]
for each $z$ and $k$. Its primitives are potential outcomes and potential treatment decisions, not a comparison of the observed treated and untreated groups. The quantity $\omega^d_{zk}$ is the relative treatment-selection odds associated with potential category $k$, using category $1$ as the reference category. It measures how the odds of treatment selection differ for individuals whose potential outcome is category $k$, relative to individuals whose potential outcome is category $1$, conditional on $Z=z$. Thus, $\omega^d_{zk}>1$ indicates relatively stronger selection into treatment for category $k$, while $\omega^d_{zk}<1$ indicates relatively weaker selection. The choice of reference category is arbitrary. Changing the reference category changes the numerical values of the individual odds ratios, but it does not change the content of Assumption~\ref{a:similar}. For each $z$, association similarity equates the category-specific odds ratios in the two potential-outcome tables. Using category $1$ as the reference, this restriction is
\[
\frac{q^0_{z,1,0}q^0_{z,k,1}}
     {q^0_{z,1,1}q^0_{z,k,0}}
=
\frac{q^1_{z,1,0}q^1_{z,k,1}}
     {q^1_{z,1,1}q^1_{z,k,0}},
\qquad k=2,\ldots,q.
\]
Rearranging gives
\[
\frac{q^0_{z,k,1}q^1_{z,k,0}}
     {q^0_{z,k,0}q^1_{z,k,1}}
=
\frac{q^0_{z,1,1}q^1_{z,1,0}}
     {q^0_{z,1,0}q^1_{z,1,1}}.
\]
Consequently, the ratio of the missing-cell product to the observed-cell product is constant across categories. Since all cells are positive, write this common ratio as $e^{u_z}$. This gives
\begin{equation}\label{eq:product}
q^0_{z,k,1}q^1_{z,k,0}
=
e^{u_z}q^0_{z,k,0}q^1_{z,k,1},
\qquad k=1,\ldots,q.
\end{equation}
The scalar $u_z$ is defined by
\[
u_z
=
\log\!\left(
\frac{q^0_{z,k,1}/q^0_{z,k,0}}
     {q^1_{z,k,1}/q^1_{z,k,0}}
\right),
\]
which is the same for every $k$ under association similarity. It measures the difference in log treatment-selection odds when category membership is defined using $Y_0$ rather than $Y_1$. Association similarity allows those odds to vary across categories, but requires this difference to remain constant. Introducing $u_z$ adds no restriction: it simply expresses association similarity through a common positive multiplier. There is one unknown multiplier for each instrument value, so the remaining unknown pair is $u=(u_0,u_1)$. Once the missing cells at $z=0$ are known, the missing cells at $z=1$ follow from (\ref{eq:dominance}). Fix a trial pair $(u_0,u_1)$. For each category, equation (\ref{eq:product}) at the two instrument values becomes
\begin{equation}\label{eq:root}
\begin{aligned}
q^0_{0,k,1}(u)q^1_{0,k,0}(u)
&=
e^{u_0}q^0_{0,k,0}q^1_{0,k,1},
\\
\bigl[q^0_{0,k,1}(u)+\Delta q^0_k\bigr]
\bigl[q^1_{0,k,0}(u)-\Delta q^1_k\bigr]
&=
e^{u_1}q^0_{1,k,0}q^1_{1,k,1}.
\end{aligned}
\end{equation}
For a fixed $u$, everything on the right-hand sides is known. The two unknowns are the missing cells $q^0_{0,k,1}(u)$ and $q^1_{0,k,0}(u)$. The first equation expresses one missing cell in terms of the other. Substituting into the second produces a quadratic equation.

Equation (\ref{eq:root}) by itself does not determine $u$. For any trial value of $u$, it gives a category-by-category candidate completion of the latent tables. The value of $u$ must be selected by requiring the completed columns to have the correct probability totals.

At this stage, therefore, the recovered entries need not sum to the required column totals. The missing $Y_0$ column at $z=0$ describes people with $D_0=1$, so its entries must sum to $p_0$. The missing $Y_1$ column describes people with $D_0=0$, so its entries must sum to $1-p_0$. Hence $u$ must satisfy
\begin{equation}\label{eq:F}
\sum_k q^0_{0,k,1}(u)=p_0,
\qquad
\sum_k q^1_{0,k,0}(u)=1-p_0.
\end{equation}
These are two equations in the two unknowns $u_0$ and $u_1$. They select the unique value of $u$ for which the category-wise solutions obtained from (\ref{eq:root}) form valid probability columns.

The following condition is not a primitive behavioral assumption. It is a testable condition on the observed distribution that is sufficient for point identification. Its role is analogous to that of a monotone likelihood ratio condition in IVQR analysis: it is a monotonicity condition used to obtain uniqueness and invertibility of the identification equations.

\begin{condition}[Testable categorywise dominance]
\label{a:dominance}
For every category $k$,
\[
\Delta q^0_k>0,
\qquad
\Delta q^1_k>0.
\]
\end{condition}

Because $q^0_{0,k,0}$, $q^0_{1,k,0}$, $q^1_{1,k,1}$, and $q^1_{0,k,1}$ are identified from the observed data, Condition~\ref{a:dominance} can be checked in the population and through its sample analogue in empirical work. It is stronger than aggregate first-stage relevance. It requires every observed category probability to move in the same direction when the instrument changes. If the condition fails, the point-identification argument given here does not apply, although partial identification may still be possible.

\begin{theorem}[Global identification]
\label{t:global}
Suppose Assumptions~\ref{a:iv}, \ref{a:positive}, and \ref{a:similar} hold. Then the potential outcome distribution is identified if the categorywise dominance condition~\ref{a:dominance} is satisfied.
\end{theorem}

\subsection{Sensitivity and partial identification}

Point identification requires association similarity. I now allow this restriction to fail and examine how the identified treatment effects change. For each instrument value $z$ and category $k$, define
\begin{equation}\label{eq:h}
h_{zk}
=
\log\omega^0_{zk}-\log\omega^1_{zk},
\qquad h_{z1}=0.
\end{equation}
The quantity $h_{zk}$ measures the difference between the two log odds
ratios, using category 1 as the reference. Association similarity requires
$h_{zk}=0$ for every $z$ and $k$. Allowing a departure $h$ changes the product equations to
\[
q^0_{z,k,1}q^1_{z,k,0}
=
e^{u_z+h_{zk}}q^0_{z,k,0}q^1_{z,k,1}.
\]
Thus, fixing $h$ multiplies each observed-cell product by $e^{h_{zk}}$.
For any candidate $u=(u_0,u_1)$, strict categorywise dominance gives a unique admissible solution for the missing cells. Write these cells at
$z=0$ as $q^0_{0,k,1}(u;h)$ and $q^1_{0,k,0}(u;h)$. The recovered cells must also add up to the observed treatment shares.
Let $p_0=\Prb(D=1\mid Z=0)$ and define
\begin{equation}\label{eq:Fh}
F(u;h)
=
\begin{pmatrix}
\displaystyle\sum_{k=1}^{q}q^0_{0,k,1}(u;h)-p_0\\[6pt]
\displaystyle\sum_{k=1}^{q}q^1_{0,k,0}(u;h)-(1-p_0)
\end{pmatrix}.
\end{equation}
The first component measures whether the recovered insured column sums
to $p_0$. The second measures whether the recovered uninsured column
sums to $1-p_0$. Solving $F(u;h)=0$ makes both potential-outcome
distributions sum to one. For each fixed $h$, the same uniqueness argument applies. If a root
$u(h)$ exists, the potential-outcome probabilities are
\[
\pi_{0k}(h)
=
q^0_{0,k,0}+q^0_{0,k,1}(u(h);h),
\qquad
\pi_{1k}(h)
=
q^1_{0,k,1}+q^1_{0,k,0}(u(h);h).
\]
Write $\pi(h)=(\pi_0(h),\pi_1(h))$. If no root exists, that departure
$h$ is incompatible with the observed table and the maintained model. To limit the size of the departures, consider
\begin{equation}\label{eq:H}
\aset_K
=
\left\{
h:
h_{z1}=0,\quad
\max_k h_{zk}-\min_k h_{zk}\leq K,\quad z=0,1
\right\},
\qquad K\geq0.
\end{equation}
This restriction bounds $|h_{zk}-h_{z\ell}|$ by $K$ for every pair of
categories $k,\ell$. It therefore limits how much each pairwise log
odds ratio can differ across the two potential treatment states.

The bound does not depend on the reference category. Changing the
reference subtracts the same constant from every entry in a row of
$h$. This leaves the row's range unchanged, and the constant is
absorbed by $u_z$. At $K=0$, all departures are zero, so I recover
association similarity and, when the model is compatible with the
data, the point-identified value.

One could instead impose a directional restriction, such as
$h_{zk}\geq0$. Its meaning depends on the chosen reference category.
For nominal outcomes, there is no natural category ordering that
determines such a direction.

\begin{proposition}[Sharp sensitivity set]
\label{p:sharp}
Suppose Assumptions~\ref{a:iv}, \ref{a:positive}, and
\ref{a:dominance} hold. For a specified set $\aset$ of allowable
departures, the sharp identified set for the two potential-outcome
marginals is
\begin{equation}\label{eq:set}
\mathcal I(\aset)
=
\left\{
\pi(h):
h\in\aset,\quad
\exists u\in\R^2 \text{ such that } F(u;h)=0
\right\}.
\end{equation}
For category $k$, the sharp lower and upper effect bounds are the
infimum and supremum of
\[
\pi_{1k}(h)-\pi_{0k}(h)
\]
over this set. Allowing more departures can only enlarge the
identified set.
\end{proposition}

Sharpness means that the set includes every pair of marginals allowed
by the assumptions, and that every pair in the set can actually arise
from a compatible latent joint distribution. The appendix establishes
this second claim by constructing such a distribution. The construction
has no defiers, but this is a way to prove attainability, not an
additional assumption about the actual population.

The set $\mathcal I(\aset)$ describes the two potential-outcome
distributions jointly. Separate category bounds are projections of this
set: their endpoints generally cannot be combined arbitrarily, because
the probabilities must remain jointly compatible and sum to one.
Likewise, evaluating a few selected departures gives a sensitivity
illustration, not the sharp bounds over all of $\aset_K$.

\section{Discussion of association similarity}

\subsection{Economic interpretation and a behavioral foundation}

How should association similarity be understood economically? Consider the health insurance example. Fix the instrument and pick two health categories, say good and fair. The assumption says the following: the odds of enrolling in insurance for someone whose health would be good relative to someone whose health would be fair must be the same whether I evaluate health under insurance, $Y_1$, or without insurance, $Y_0$.

Crucially, the people in a given category need not be the same across the two evaluations. Insurance may shift individuals between categories. What similarity restricts is not the enrollment rule itself, but the way treatment relates to the composition of categories in a manner that would alter selection patterns. It is a restriction on the stability of sorting. It permits insurance to improve health, to generate adverse selection, and to produce heterogeneous effects across individuals. What it rules out is a specific form of interaction: the possibility that the link between unobserved enrollment propensity and category membership changes when treatment status changes.

A simple behavioral model makes the content transparent. Suppose each individual carries a persistent risk type $T$: a diagnosis, a chronic condition, or a combination of traits that determines both insurance demand and baseline health risk. The individual enrolls when the net benefit exceeds an idiosyncratic cost:
\begin{equation}\label{behaviorchoice}
D_z=\mathbf{1}\{b(T)+s_z(T)-\varepsilon\geq 0\},
\end{equation}
where $b(T)$ captures expected medical benefits, $s_z(T)$ is the employer contribution under instrument value $z$, and $\varepsilon$ is a logistic enrollment cost. The key feature is that the same cost draw $\varepsilon$ governs both potential treatment decisions $D_0$ and $D_1$.

Health outcomes are generated by
\[
Y_d=g_d(T,\eta_d),
\]
where $\eta_d$ is a subsequent health shock. Within a risk type, enrollment costs are independent of health shocks. Formally, the categories partition into disjoint sets $K_t$ such that
\[
\Pr(Y_d\in K_t\mid T=t)=1,
\qquad d=0,1.
\]
Insurance can redistribute health within each risk group: improving symptom control, reducing pain, or shifting someone from fair to good. But it does not change the underlying diagnosis or risk profile that drives enrollment. In the insurance application, a natural reading is that insurance improves management of a chronic condition without removing the condition itself.

\begin{proposition}[A sufficient behavioral model]
\label{p:behavior}
Suppose (\ref{behaviorchoice}) holds with $\varepsilon$ having a positive density, the type-preservation condition above is satisfied, and $Z$ is independent of $(T,\varepsilon,\eta_0,\eta_1)$ and affects only $s_z(T)$. Then association similarity holds at both instrument values.
\end{proposition}

\noindent\textit{Proof.} For any category $c_k\in K_t$, observing that category reveals $T=t$. Conditional independence of $\varepsilon$ and $(\eta_0,\eta_1)$ given $T$ implies
\[
\Pr(D_z=1\mid Y_d=c_k)
=
\Lambda\{b(t)+s_z(t)\},
\qquad d=0,1,
\]
where $\Lambda$ is the CDF of $\varepsilon$. These selection probabilities do not depend on $d$, so neither do their pairwise odds ratios across categories. They may vary across categories through $t$, generating endogenous selection.
\hfill$\square$

The proposition establishes an economically interpretable setting in which association similarity holds. It is sufficient but not necessary. Indeed, the behavioral model implies the stronger equality
\[
\Pr(D_z=1\mid Y_0=c_k)=\Pr(D_z=1\mid Y_1=c_k),
\]
whereas association similarity is weaker.

\subsection{Connection with rank similarity}

The identifying role of association similarity is best understood by analogy with rank similarity in the instrumental quantile model of \citet{chernozhukovhansen2005}. In that framework, a continuously distributed outcome is written as
\[
Y_d=Q_d(U_d),
\]
where $Q_d$ is an increasing quantile function and $U_d$ is a latent rank uniformly distributed on $[0,1]$. Treatment selection is governed by some variable $V$. Rank similarity requires that, conditional on $(V,Z)$, the distribution of $U_0$ equals the distribution of $U_1$. In words, once I condition on the information that drives selection, the distribution of latent rank is the same regardless of treatment status. The quantile functions $Q_0$ and $Q_1$ may differ freely, so treatment can have arbitrary effects on the outcome scale. What is held fixed is the rank distribution. Rank invariance, the stronger condition $U_0=U_1$ almost surely, pins down individual ranks across treatment states.

Association similarity plays an analogous role in the categorical setting. It holds a dependence object fixed across potential treatment states while permitting the outcome margins $\pi_0$ and $\pi_1$ to differ. The standardized object is now the odds-ratio structure of the joint distribution of $(Y_d,D_z)$ rather than a conditional rank distribution. The analogy is one of motivation and identifying strategy, not mathematical equivalence.

There are three important differences. First, the objects being held fixed are fundamentally different. Rank similarity restricts a conditional distribution of a continuous latent variable. Association similarity restricts a finite set of odds ratios in a probability table. Neither implies the other. Second, coarsening a rank-invariant continuous model does not automatically produce association similarity. Suppose $Y_0$ and $Y_1$ are generated by a rank-invariant model and then coarsened at fixed thresholds into categorical outcomes. The rank intervals corresponding to each category can shift across treatment states because the quantile functions $Q_0$ and $Q_1$ differ. This changes the categorical selection odds ratios even when the underlying ranks are identical. Third, equality of finite-table odds ratios supplies no latent-rank representation and imposes no restrictions on within-category ranks. For nominal outcomes, such ranks are not intrinsically defined.

\section{Estimation}
\label{sec5}

The estimator replaces the population probabilities in the identification formulas with sample frequencies. It estimates the joint outcome--treatment probabilities separately within each instrument group. Here is the algorithm for the estimation:

\begin{center}
\textbf{Algorithm: estimation of the population potential-outcome distributions}
\begin{enumerate}
\item \textbf{Estimate the observed tables.} Compute the sample cells $\widehat q^d_{z,k,d}$ and the treatment rates $\widehat p_z$ within each instrument group. For an independent sample $(Y_i,D_i,Z_i)_{i=1}^n$, let
\[
n_z=\sum_i \mathbb 1\{Z_i=z\}.
\]
The observed cells and treatment rates are estimated by
\[
\widehat q^d_{z,k,d}
=
\frac{1}{n_z}
\sum_{i=1}^n
\mathbb 1\{Z_i=z,D_i=d,Y_i=c_k\},
\qquad
\widehat p_z
=
\sum_k \widehat q^1_{z,k,1}.
\]

\item \textbf{Recover the missing cells for a trial pair $(u_0,u_1)$.} Insert the sample cells into the sample analogue of (\ref{eq:root}). For each category $k$, compute the unique positive quadratic root given in the proof of Theorem~\ref{t:global}. This yields the candidate missing cells
\[
\widehat q^0_{0,k,1}(u)
\quad\text{and}\quad
\widehat q^1_{0,k,0}(u).
\]
No separate numerical optimization is required at the category level.

\item \textbf{Solve for the unknown pair $\widehat u$.} Find $\widehat u=(\widehat u_0,\widehat u_1)$ satisfying the sample normalization equations
\[
\sum_k \widehat q^0_{0,k,1}(\widehat u)=\widehat p_0,
\qquad
\sum_k \widehat q^1_{0,k,0}(\widehat u)=1-\widehat p_0.
\]
The implementation can start $(0,0)$ and can use the analytical Jacobian given in the appendix.

\item \textbf{Construct the estimated potential-outcome distributions.} Add the observed and recovered cells as in (\ref{eq:identify}):
\[
\widehat\pi_{0k}
=
\widehat q^0_{0,k,0}
+
\widehat q^0_{0,k,1}(\widehat u),
\qquad
\widehat\pi_{1k}
=
\widehat q^1_{0,k,1}
+
\widehat q^1_{0,k,0}(\widehat u).
\]
The estimated category effects are
\[
\widehat\tau_k
=
\widehat\pi_{1k}-\widehat\pi_{0k}.
\]
\end{enumerate}
\end{center}

Thus, regardless of the number of categories, only two unknowns, $u_0$ and $u_1$, are solved numerically. Each estimated potential-outcome distribution sums to one, and the estimated effects sum to zero:
\[
\sum_k \widehat\pi_{0k}=1,
\qquad
\sum_k \widehat\pi_{1k}=1,
\qquad
\sum_k \widehat\tau_k=0.
\]

Under the interior conditions, the implicit function theorem implies that the recovered margins are a continuously differentiable function of the observed probabilities in a neighborhood of the population distribution. For independent and identically distributed observations and fixed $q$, standard delta-method and bootstrap results therefore apply; see \cite{vanderVaart1998}. No separate inference result is needed for the identification argument itself. In theory, one can also use covariates; I develop that idea in the appendix.

\section{Application: Private insurance and health}

Using the replication data of \citet{acerenza2023testing}, the application studies how private insurance changes the distribution of self-reported physical and mental health. It is the 2010 Medical Expenditure Panel Survey, restricted to working adults aged 25--64 without public insurance and excluding self-employed and unemployed respondents. For each outcome, define three mutually exclusive and exhaustive categories:
\[
\begin{aligned}
c_1&=\text{excellent health},\\
c_2&=\text{good health},\\
c_3&=\text{poor health}.
\end{aligned}
\]
Each respondent belongs to exactly one category for each health measure. I define the binary instrument directly from employer size:
\[
Z=\begin{cases}
1,&\text{employer has at least 50 employees},\\
0,&\text{employer has fewer than 50 employees}.
\end{cases}
\]
This cutoff also equals the sample median employee count and gives reasonably balanced groups. Treatment is private insurance. No covariates or weights enter estimation.

There are 46.7\% respondents at $Z=0$ and 53.3\% at $Z=1$. Their insurance rates are 52.28\% and 77.41\%, respectively, giving a first stage of 25.13 percentage points. All 12 category-wise dominance differences are strictly positive in the observed sample. I solve the association-similarity equations directly on the three-category tables.

\begin{table}[htbp]
\centering\small
\setlength{\tabcolsep}{4pt}
\caption{Insurance effects in percentage points}\label{tab:insurance_three}
\begin{tabular}{llrrrr}
\hline
Outcome & Category & Effect & SE & 95\% CI  \\
\hline
Physical & Excellent/very good & 7.25 & 4.27 & [-1.01, 15.73]   \\
Physical & Good & -4.96 & 3.92 & [-12.89, 2.50]   \\
Physical & Fair/poor & -2.29 & 2.51 & [-7.94, 1.96]  \\
Mental & Excellent/very good & 8.51 & 4.64 & [0.93, 20.80]   \\
Mental & Good & -5.05 & 3.62 & [-12.91, 2.51]  \\
Mental & Fair/poor & -3.46 & 4.18 & [-18.01, -0.17]  \\
\hline
\end{tabular}
\end{table}

Table~\ref{tab:insurance_three} reports the estimated effect of private insurance on the probability of being in each health category, expressed in percentage points. Since the three categories are mutually exclusive and exhaustive, the estimated effects within each outcome sum to zero. Thus, a positive effect in one category necessarily corresponds to a reallocation of probability mass away from the other categories.

For physical health, the point estimates suggest a favorable shift in the distribution: private insurance is estimated to increase the probability of reporting excellent or very good health by 7.25 percentage points, while reducing the probabilities of good health and fair/poor health by 4.96 and 2.29 percentage points, respectively. However, all three 95\% confidence intervals for physical health include zero. Therefore, the data do not provide statistically significant evidence that private insurance affects the physical-health distribution in this sample.

For mental health, the estimated shift is similar in direction but somewhat stronger. The probability of reporting excellent or very good mental health increases by 8.51 percentage points, while the probabilities of good mental health and fair/poor mental health decrease by 5.05 and 3.46 percentage points, respectively. The 95\% confidence interval for excellent/very good mental health, $[0.93,20.80]$, and for fair/poor mental health, $[-18.01,-0.17]$, exclude zero. These results provide some evidence that private insurance improves the mental-health distribution by shifting individuals toward the best category and away from the worst category. The effect on the middle category, good mental health, is not statistically distinguishable from zero.

Overall, the results do not show a statistically significant effect of private insurance on physical health, but they provide category-specific evidence of a favorable effect on mental health.

\section{Conclusion}

This paper develops a nonparametric framework for identifying treatment effects when
outcomes are categorical, extending the logic of instrumental variable methods beyond
continuous and ordered settings. The starting observation is that researchers confronted
with categorical outcomes routinely either discard distributional information by binarizing
the outcome or impose restrictive parametric assumptions to recover it. This paper offers
an alternative that retains full distributional content without parametric restrictions on
the unobservables. The central contribution is the association similarity assumption, which, combined with standard IV conditions and a discrete dependence restriction, yields point identification of
the full distribution of potential outcomes for genuinely unordered categorical outcomes
using only a binary instrument. A partial identification result complements the main theorem by characterizing the identified set and providing a transparent and formally sharp foundation for sensitivity
analysis.

\bibliographystyle{chicago}
\bibliography{references}

\appendix
\section{Proofs}
\label{app:proofs}

\subsection{Proof of Theorem~\ref{t:global}}

Throughout the proof, fix the observed cells and suppose that
Condition~\ref{a:dominance} holds. Write
\[
x_k(u)=q^0_{0,k,1}(u),\qquad
y_k(u)=q^1_{0,k,0}(u),\qquad
M_{zk}(u)=e^{u_z}q^0_{z,k,0}q^1_{z,k,1}.
\]
These are shorthand for the missing cells at $z=0$ and the right-hand
sides of the product equations. In this proof, $F(u)$ means $F(u;0)$
as defined in (\ref{eq:Fh}):
\begin{equation}\label{eq:appF}
F(u)=
\begin{pmatrix}
\sum_k x_k(u)-p_0\\
\sum_k y_k(u)-(1-p_0)
\end{pmatrix}.
\end{equation}
Thus, $F(u)=0$ is exactly the pair of normalization equations in
(\ref{eq:F}). I establish existence and uniqueness of the missing
cells for each trial $u$, uniqueness of a normalization root, and
compatibility of the recovered tables with a common joint distribution.

\paragraph{Step 1: Recovering the missing cells.}
For a fixed $u\in\R^2$, equation (\ref{eq:root}) becomes
\[
x_ky_k=M_{0k},\qquad
(x_k+\Delta q^0_k)(y_k-\Delta q^1_k)=M_{1k}.
\]
Eliminating $y_k$ gives
\begin{equation}\label{eq:appquadratic}
\Delta q^1_k x_k^2-R_kx_k-\Delta q^0_kM_{0k}=0,
\qquad
R_k=M_{0k}-M_{1k}-\Delta q^0_k\Delta q^1_k.
\end{equation}
The leading coefficient is positive and the constant term is negative.
Consequently, the quadratic has exactly one positive root:
\begin{equation}\label{eq:approot}
x_k(u)=
\frac{R_k+\sqrt{R_k^2+4\Delta q^0_k\Delta q^1_kM_{0k}}}
     {2\Delta q^1_k},
\qquad y_k(u)=\frac{M_{0k}}{x_k(u)}.
\end{equation}
When $R_k<0$, the numerically equivalent expression
\[
x_k(u)=
\frac{2\Delta q^0_kM_{0k}}
     {\sqrt{R_k^2+4\Delta q^0_k\Delta q^1_kM_{0k}}-R_k}
\]
avoids subtracting nearly equal numbers.

Since $M_{1k}>0$ and $x_k+\Delta q^0_k>0$, the second product equation
implies $y_k>\Delta q^1_k$. Hence all four missing cells,
\[
x_k,\quad y_k,\quad x_k+\Delta q^0_k,\quad y_k-\Delta q^1_k,
\]
are strictly positive. They are continuously differentiable in $u$
on all of $\R^2$. Before normalization they are candidate cell masses;
they need not yet sum to their required probability totals.

\paragraph{Step 2: The Jacobian and global uniqueness.}
Suppress the argument $u$ and define
\[
J_k=y_k\Delta q^0_k+x_k\Delta q^1_k>0.
\]
Differentiating the product equations yields
\[
\begin{pmatrix}
y_k&x_k\\
y_k-\Delta q^1_k&x_k+\Delta q^0_k
\end{pmatrix}
\begin{pmatrix}dx_k\\dy_k\end{pmatrix}
=
\begin{pmatrix}M_{0k}\,du_0\\M_{1k}\,du_1\end{pmatrix}.
\]
The coefficient matrix has determinant $J_k$. Therefore the analytical
Jacobian used in estimation is
\begin{equation}\label{eq:appjacobian}
F_u(u)=
\begin{pmatrix}
\displaystyle\sum_k\frac{(x_k+\Delta q^0_k)M_{0k}}{J_k}
&-\displaystyle\sum_k\frac{x_kM_{1k}}{J_k}\\[7pt]
-\displaystyle\sum_k\frac{(y_k-\Delta q^1_k)M_{0k}}{J_k}
&\displaystyle\sum_k\frac{y_kM_{1k}}{J_k}
\end{pmatrix}.
\end{equation}
Its diagonal entries are positive and its off-diagonal entries are
negative. Using the two product equations, the diagonal-dominance
differences simplify to
\begin{equation}\label{eq:dd}
\begin{aligned}
F_{1,u_0}-|F_{1,u_1}|
&=\sum_k\frac{x_k(x_k+\Delta q^0_k)\Delta q^1_k}{J_k}>0,\\
F_{2,u_1}-|F_{2,u_0}|
&=\sum_k\frac{y_k(y_k-\Delta q^1_k)\Delta q^0_k}{J_k}>0.
\end{aligned}
\end{equation}
These inequalities hold for every $u\in\R^2$.

To establish global uniqueness, take any $u,u'\in\R^2$. The fundamental
theorem of calculus gives
\[
F(u)-F(u')=\overline J(u,u')(u-u'),\qquad
\overline J(u,u')=\int_0^1F_u\bigl(u'+t(u-u')\bigr)\,dt.
\]
Because the signs of the Jacobian entries are fixed, integration
preserves the strict inequalities in (\ref{eq:dd}). Thus
$\overline J(u,u')$ is strictly diagonally dominant and nonsingular.
Indeed, if $\overline Jv=0$ for a nonzero $v$, choosing a component
$v_i$ of largest absolute value would give
\[
|\overline J_{ii}|\,|v_i|
\leq\sum_{j\ne i}|\overline J_{ij}|\,|v_j|
\leq\sum_{j\ne i}|\overline J_{ij}|\,|v_i|,
\]
contradicting strict diagonal dominance. It follows that
$F(u)=F(u')$ implies $u=u'$. In particular, $F$ has at most one root.

\paragraph{Step 3: Existence under the maintained model.}
Any population satisfying Assumptions~\ref{a:iv}, \ref{a:positive},
and \ref{a:similar} supplies positive missing cells and finite
multipliers $u_0,u_1$ satisfying (\ref{eq:product}). Its common
potential-outcome margins imply (\ref{eq:root}), and its treatment
shares imply $F(u)=0$. Hence a root exists under the maintained model.
Combining this fact with Step 2 identifies $u$ and the two margins:
\begin{equation}\label{eq:identify}
\pi_{0k}=q^0_{0,k,0}+x_k(u),\qquad
\pi_{1k}=q^1_{0,k,1}+y_k(u).
\end{equation}
This existence statement uses model compatibility; a numerical root
is not assumed to exist for an arbitrary observed table.

\paragraph{Step 4: A common joint distribution.}
At any root, consider the following unconditional outcome margins
within three treatment-response strata:
\[
\begin{array}{c|c|c|c}
(D_0,D_1)&\text{stratum mass}&Y_0=c_k&Y_1=c_k\\
\hline
(1,1)&p_0&x_k&q^1_{0,k,1}\\
(0,1)&p_1-p_0&\Delta q^0_k&\Delta q^1_k\\
(0,0)&1-p_1&q^0_{1,k,0}&y_k-\Delta q^1_k
\end{array}
\]
All entries are positive. The two margin vectors in each row have the
same total, since $F(u)=0$ and
\[
\sum_k\Delta q^0_k=\sum_k\Delta q^1_k=p_1-p_0>0.
\]
For a stratum $s$ with mass $w_s$ and margin vectors
$(a_{sk})_k$ and $(b_{s\ell})_\ell$, define
\[
\Prb(Y_0=c_k,Y_1=c_\ell,(D_0,D_1)=s)
=\frac{a_{sk}b_{s\ell}}{w_s}.
\]
Summing these masses across the three strata gives a probability law
for $(Y_0,Y_1,D_0,D_1)$. Draw $Z$ independently with its observed
distribution and set $D=D_Z$ and $Y=Y_D$.

At $z=0$, the untreated $Y_0$ masses are
$\Delta q^0_k+q^0_{1,k,0}=q^0_{0,k,0}$ and the treated $Y_1$ masses
are $q^1_{0,k,1}$. At $z=1$, the untreated $Y_0$ masses are
$q^0_{1,k,0}$ and the treated $Y_1$ masses are
$q^1_{0,k,1}+\Delta q^1_k=q^1_{1,k,1}$.
The remaining columns equal the recovered missing cells. Thus this
single law reproduces both observed tables and all completed potential
tables. Equation (\ref{eq:product}) implies association similarity.
The construction has no defiers, but no-defier behavior was not used
to derive uniqueness or required of every admissible population.
\hfill$\square$

\subsection{Proof of Proposition~\ref{p:sharp}}

Maintain Assumptions~\ref{a:iv} and \ref{a:positive}, together with
Condition~\ref{a:dominance}. For any fixed allowable departure $h$,
replace $M_{zk}$ in the preceding proof by
\[
M_{zk}(u;h)=e^{u_z+h_{zk}}q^0_{z,k,0}q^1_{z,k,1}.
\]
These quantities are strictly positive. Their derivatives with respect
to $u_z$ equal themselves, so the positive-root formula, the analytical
Jacobian, and the global uniqueness argument remain valid for each
fixed $h$. Consequently, $F(u;h)=0$ has at most one root.

\paragraph{Necessity.}
Take any joint distribution satisfying the maintained restrictions
and having departures $h\in\aset$. Its odds ratios give
\[
\frac{q^0_{z,k,1}q^1_{z,k,0}}
     {q^0_{z,k,0}q^1_{z,k,1}}
=e^{u_z+h_{zk}},
\]
where $u_z$ is the log product ratio for reference category 1.
Common potential-outcome margins give the modified category equations,
and the treatment shares give $F(u;h)=0$. Its marginals therefore
belong to $\mathcal I(\aset)$ in (\ref{eq:set}).

\paragraph{Attainability.}
Conversely, take $h\in\aset$ and a root $u$ of $F(u;h)=0$.
Apply the principal-stratum construction in Step 4 to the resulting
positive missing cells. It produces a joint distribution satisfying
instrument validity and interiority and reproducing the observed
tables. To verify its departure values, let
\[
L_{zk}=\log\left(
\frac{q^0_{z,k,1}q^1_{z,k,0}}
     {q^0_{z,k,0}q^1_{z,k,1}}\right)=u_z+h_{zk}.
\]
The odds ratios in (\ref{eq:or}) satisfy
\[
\log\omega^0_{zk}-\log\omega^1_{zk}
=L_{zk}-L_{z1}=h_{zk},
\]
because $h_{z1}=0$. Thus every element of $\mathcal I(\aset)$ is
attainable under exactly the stipulated departures. Together with
necessity, this proves sharpness.

Taking the infimum and supremum of $\pi_{1k}-\pi_{0k}$ over the joint
set gives the sharp lower and upper category-effect bounds. Endpoints
need not be attained when strict positivity excludes limiting tables.
If $\aset\subseteq\widetilde{\aset}$, then
$\mathcal I(\aset)\subseteq\mathcal I(\widetilde{\aset})$, proving
nesting.

Finally, if category $j$ becomes the reference, set
\[
h'_{zk}=h_{zk}-h_{zj},\qquad u'_z=u_z+h_{zj}.
\]
Then $u'_z+h'_{zk}=u_z+h_{zk}$, so the completed cells and margins
are unchanged. Subtracting a row constant also preserves
$\max_k h_{zk}-\min_k h_{zk}$. This proves reference invariance of
the restriction in (\ref{eq:H}).
\hfill$\square$

\section{Adjustment for observed covariates}
\label{app:covariates}

This section describes a theoretical extension. Replace the independence component of
Assumption~\ref{a:iv} by
\[
(Y_0,Y_1,D_0,D_1)\perp Z\mid X,
\qquad 0<e(X)<1\ \text{almost surely},
\]
where $e(X)=\Prb(Z=1\mid X)$. For each $d,z,k$,
\begin{align*}
\E_X\Prb(Y=c_k,D=d\mid Z=z,X)
&=\E_X\Prb(Y_d=c_k,D_z=d\mid X)\\
&=\Prb(Y_d=c_k,D_z=d)=q^d_{z,k,d}.
\end{align*}
Thus, standardizing each observed table to the same covariate distribution identifies the observed columns of the marginal potential tables.

Apply the identification equations to these standardized cells with the same restrictions on the resulting conditional tables. Association similarity conditional on $X$ is not a substitute: equality of conditional odds ratios need not imply equality after averaging over $X$. Likewise, feasibility of
the standardized tables alone does not establish compatibility with every feature of the observed conditional distribution.

For estimation, write $e_1(x)=e(x)$ and $e_0(x)=1-e(x)$. A normalized
inverse-propensity estimator is
\begin{equation}\label{eq:appipw}
\widehat q^d_{z,k,d}=
\frac{\displaystyle\sum_{i=1}^n
 \frac{\mathbb 1\{Z_i=z,D_i=d,Y_i=c_k\}}{\widehat e_z(X_i)}}
 {\displaystyle\sum_{i=1}^n
 \frac{\mathbb 1\{Z_i=z\}}{\widehat e_z(X_i)}}.
\end{equation}
The common denominator makes the cells sum to one within each
instrument group. Set
$\widehat p_z=\sum_k\widehat q^1_{z,k,1}$ and use these cells and treatment shares in the same two-equation algorithm.

\end{document}